# Phase estimation with autoregressive padding (PEAP): addressing inaccuracies and biases in EEG analysis

Miriam Kirchhoff, Johanna Rösch, Maria Ermolova, Oskari Ahola, Sarah Harders, Juliana Hougland, Ulf Ziemann

Accurate phase estimation at the edge of data segments is crucial for EEG applications such as EEG-TMS in offline and real-time data analysis. Our research evaluates the phase estimation performance of four commonly used methods (Phastimate, SSPE, ETP, and PhastPadding) for accuracy and systemic biases, using data from young and elderly healthy controls and chronic stroke participants. To address the identified limitations of the established methods, we introduce Phase Estimation with Autoregressive Padding (PEAP), a method that prevents strong bandpass filtering-induced artifacts.

Contrary to the established methods, PEAP does not show significant biases and improves accuracy by 3.2 to 9.2 % for the continuous phase estimation. Our offline analysis demonstrates how established methods are systematically biased towards some estimates and how they induce phase shifts. We also show that differences between methods do not vary between clinical and control populations, supporting their translatability.

This work indicates that systematic biases in established phase estimation methods may compromise the validity and comparability of phase-dependent findings. PEAP addresses these limitations and thus offers a more reliable and more accurate alternative method.

## Introduction

Estimating the instantaneous phase of ongoing electroencephalographic (EEG) oscillations provides insight into underlying neural processes. For example, the phase of the rolandic mu rhythm (8-13 Hz) has shown to be indicative of corticospinal excitability in offline analyses of EEG signals during transcranial magnetic stimulation (TMS) of the primary motor cortex (Zrenner et al., 2022). This relationship has been translated into real-time, phase-dependent stimulation paradigms in healthy participants (Humaidan et al., 2024; Kirchhoff et al., 2024; Zrenner et al., 2018) and stroke patients (Wischnewski et al., 2025). For such offline analyses and real-time applications, accurate phase estimation is crucial (Zrenner et al., 2020).

However, estimating the instantaneous phase, i.e., the phase of the last sample in a signal or epoch, poses a methodological challenge. Standard methods to obtain phase information from a continuous signal, which extract the target oscillation using an acausal filter followed by a Hilbert transform, do not suffice for the instantaneous phase. For the last sample in a signal, both filtering and the Hilbert transform induce edge artifacts that distort the phase (Rogasch et al., 2017), leading to systematic biases and inaccuracies in the phase estimate. Additionally, real-time applications often require forecasting which phase will occur. Consequently, specialized methods are needed for reliable instantaneous phase estimation.

Several methods have been proposed for instantaneous phase estimation in both offline analyses and in real-time, phase-dependent stimulation (Chang et al., 2025; Shirinpour et al., 2020; Wodeyar et al., 2021; Zrenner et al., 2020). Several conventional methods, such as Phastimate (Zrenner et al., 2020) and Educated Temporal Prediction (ETP) (Shirinpour et al., 2020), filter the data and then interpolate the instantaneous phase from earlier signals that are less affected by the filtering edge artifact. Phastimate uses autoregressive models (AR) for interpolation, while ETP assumes a sinusoidal signal and calculates the instantaneous phase based on the last detected peak. An alternative approach proposed by Wodeyar et al. (2021) employs a state-space model to estimate the current phase, bypassing the need for bandpass filtering altogether. To date, there is no consensus on which method to use and in how far their estimates differ systematically.

The established methods described above either attempt to correct the filtering edge artifact or avoid filtering entirely. Here, we propose an alternative strategy: preventing the filtering edge artifact through padding of the unfiltered data. Phase Estimation through Autoregressive Padding (PEAP) extends the unfiltered signal with simulated, oscillatory data to shift the edge artifact away from the target sample. Unlike traditional padding methods, such as zero-padding or reflect padding, this strategy preserves the continuity of the ongoing oscillation, minimizing phase distortion (Chang et al., 2025). PEAP does not filter prior to padding, unlike Phastimate-based padding (Chang et al., 2025). By avoiding truncation of real data and moving the filtering edge artifact, we expect this method to increase phase estimation accuracy and reduce biases. A related approach has been proposed recently and demonstrated to improve accuracy. However, the work by Chang et al. (2025) suffers from poor interpretability since it includes several artifact-inducing steps that might enhance each other, like filtering the data several times.

We compare PEAP to established methods of instantaneous phase estimation using EEG signals from healthy controls and chronic stroke patients. We provide an overview on the accuracy and characteristic biases of the different methods, as well as insights on how these may affect analyses.

## Methods

### Datasets

The preexisting dataset we used consists of in total 72 subjects drawn from three data sub-sets:

1. Healthy young control subjects: 54 subjects (34 f, 20 m), mean age 25.2 years ($sd$ = 4.1). Parts of this dataset have been previously published in Kirchhoff et al. (2024) and Kirchhoff et al. (2026).
2. Healthy elderly control subjects: 10 subjects (5 f, 5 m), mean age 54.4 years ($sd$ = 8.9).
3. Stroke patients: 8 subjects (2 f, 4 m), mean age 59.0 years ($sd$ = 9.6), average time post stroke in months: 71.3 ($sd$ = 61.7)).

Healthy participants were excluded from participation if they could not consent, had any contraindications based on the guidelines on safety of TMS application (Rossi et al., 2021) , or contraindications to magnetic-resonance imaging (MRI) scans, additionally if they had any known history of head trauma or other study-relevant pre-existing conditions.

Stroke patients were recruited if they complied with the following inclusion criteria: 1) age of 18-85 years, 2) chronic stroke ($\geq$ 6 months post-stroke) with persisting hand/arm paresis and spasticity, 3) ability to evoke ipsilesional MEPs, and 4) be able to consent to the protocol. Exclusion criteria were 1) a history of seizures, 2) taking pro-convulsive or muscle relaxing medication, or 3) having other contraindications based on the guidelines on safety of TMS application (Rossi et al., 2021) or 4) contraindications to MRI scans. Individual datasets were included only if the resting-state EEG retained for data analysis was longer than 4 min (1 dataset could not be included due to this criterion).

For all populations, the data were recorded using a 128-channel EEG system according to the international 10-5 system (EasyCap BC-TMS-128, EasyCap, Herrsching, Germany). We kept channel impedances below 5 k$\Omega$. 5-10 minutes of resting-state EEG with open eyes were recorded at a sampling rate of 5 kHz with a Bittium NeurOne Tesla EEG System including amplifiers (Bittium Corporation, Oulu, Finland).

All participants signed informed consent, were informed about the study purpose and the voluntary nature of participation and received monetary reimbursement. The study was approved by the ethics committee at the medical faculty of the University of Tübingen (young healthy: project 810/2021BO2, elderly healthy: project 530/2019BO1, stroke patients: project 715/2021BO) and conducted in accordance with the revised declaration of Helsinki (World Medical Association, 2013).

## EEG processing and phase extraction

### Train-test split

Our dataset consisted of a total 72 resting-state EEG signals. These recordings were split into a training set (first 3 min, following Shirinpour et al. (2020)) and a testing set (remaining signal). To rule out any data leak, all following analyses were conducted on the training and test set separately.

### Data preparation

Markers were generated in the resting-state EEG with random inter-stimulus intervals (ISI; 1.25 - 1.75 s), reflecting the time of virtual TMS pulses. To calculate the ground truth, we used the continuous signal. Otherwise, we epoched the sample at [-2065 -1] ms with respect to the generated TMS markers. The long epoch was selected since one of the models (state-space phase estimate; SSPE) requires at least 2 s epoch length. If epochs contained channels with an EEG amplitude range above 150 µV, they were automatically rejected under the assumption that they contained noise (see Zrenner et al., 2022, Kirchhoff et al., 2024).

Epoched and continuous data were downsampled from 5000 Hz to 1000 Hz and detrended linearly using the TESA toolbox (Mutanen et al., 2020; Rogasch et al., 2017). We then used a Laplacian montage around C3 (surround electrodes: FCC5h, FCC3h, CCP5h, CCP3h) for all healthy subjects. For stroke subjects, we extracted the Laplacian around C3 and C4 (surround electrodes: FCC6h, FCC4h, CCP6h, CCP4h) and grouped them into the affected and intact hemisphere stroke dataset, depending on stroke location.

*Ground truth calculation*

In line with (Shirinpour et al., 2020), the ground truth was calculated separately from the continuous pre-processed training and test sets. The data were filtered using an acausal bandpass FIR filter between 9-13 Hz (order 230) with a hamming window of the same order. The phase of the resulting oscillation was extracted by a Hilbert transform.

*Padding methods and filtering*

The downsampled, detrended, and Laplace-transformed data were either not padded, padded with PEAP, or padded with Phastimate-based padding (PhastPadding, Chang et al.,(Chang et al., 2025)) before filtering. Then, the epoch was shortened to [-980 -1] ms (input length).

PEAP: For PEAP, we used a Burg autoregressive model (model type) with model order 130 to predict 290 ms (output length) of data. We trained the model on each trial separately to get realistic forecasting. The input length, output length, model type, and model order were previously determined through hyperparameter tuning.

PhastPadding: The data were filtered (for specifications, see ground truth calculation), processed through Phastimate with standard settings (order = 30, window = 128, offset = 0, edge = 65), which was used for 100 ms of forecasting. The forecasted data (t ≥ 0) was then appended to the non-filtered data.

Padded and unpadded data were then filtered through an acausal bandpass FIR filter between 9-13 Hz with order 230 and hamming window (also order 230).

*Phase extraction*

The filtered data (padded or unpadded) were used to estimate the phase at the time of the TMS marker using four approaches (see Fig. 1): (1) The data were Hilbert transformed. (2) Phastimate was applied with individualized, optimized settings (see *Hyperparameter tuning*). Phastimate cuts the edge of the data (at default, -65 ms at both ends) to remove edge artifacts, then interpolates the data beyond 0 using an autoregressive model and finally uses a Hilbert transform. (3) ETP also cuts the data to avoid the edge artifact, but at -40 ms. It then searches for the most recent peak and extrapolates the future data as a sine wave with individualized inter-peak interval (see *Hyperparameter tuning*). (4) SSPE differs from the other methods in that it does not use filtered data, but longer epochs. The algorithm uses min. 2 seconds of raw data to calculate a state-space model that infers the phase of the ongoing oscillation. We used the data of [-2064 -1] ms for the calculation.

*Hyperparameter tuning*

For PEAP, Phastimate, and ETP, parameters were optimized on the training set. To optimize PEAP, we conducted a grid search through possible hyperparameter settings on the training set. Specifically, we tested varying the input epoch length (700 - 1000 ms in steps of 20), number of samples to be interpolated ("output length", 110 - 300 ms in steps of 20), order of the AR model (110 - 200 in steps of 10) and AR model type (Burg or Yule-Walker). We evaluated the performance of approaches by means of their circular root mean squared error between ground truth and estimate. To avoid combinations that lead to biased estimates, hyperparameter combinations were excluded if a Kuiper test (Berens, 2009) between the results and ground truth was significant, i.e., when it was likely that both were sampled from different distributions. To address the trade-off between complexity and accuracy, a more complex model was only considered if it improved accuracy by at least .1 % compared to the next best model. Through this procedure, we calculated the hyperparameters across participants.

For ETP, we extracted individualized inter-peak intervals to adjust the period of the sine wave used for interpolation. For Phastimate, we individualized the optimal window length, filter order, edge length, and order of the autoregressive model.

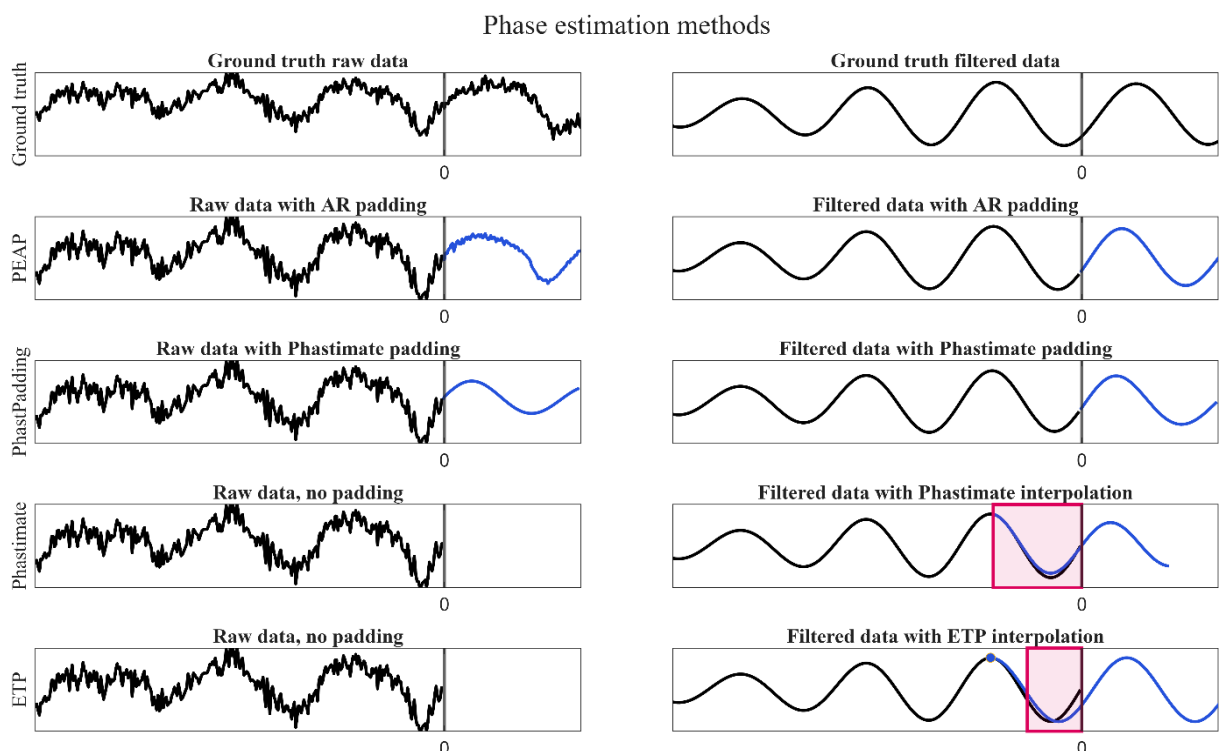

**Figure 1: Schematic overview of the phase estimation methods.** The left column illustrates the data before bandpass-filtering (black line) without or with padding (blue line). The right column shows the data after filtering (black line) and the interpolation of data (blue line) for Phastimate or ETP, along with the edge that was cut (pink area).

**First row: Ground truth.** Continuous data before filtering (left) and after filtering (right).

**Second row: PEAP.** An autoregressive model is fitted to the unfiltered trial data, which is then used to interpolate data from t = 0 on (blue line, left plot). After filtering, the edge artifact is shifted to the interpolated data. A Hilbert transform is used to determine the phase.

**Third row: PhastPadding.** Similar to PEAP, except that the interpolated data is retrieved from applying the Phastimate algorithm.

**Fourth row: Phastimate.** No padding is used for filtering. After filtering, the data that may contain edge artifacts (default: from -65 ms on) is cut. An autoregressive model is fit to the filtered trial data and used to interpolate the data from the cutoff beyond time 0. The phase is determined with a Hilbert transform.

**Fifth row: ETP.** No padding is used. The filtered data is cut at -40 ms. The last peak is found. From there, a sinusoid with individualized frequency is fitted to estimate the phase

**Data analysis**

*Quality control*

The power frequency spectrum was computed (Welch's method, window size 2s) and plotted for the full Hjorth-filtered signal of each subject for the training and test set separately.

Based on the power frequency spectrum, the spectral signal-to-noise ratio (SNR) was computed for our target frequency (9-13 Hz), with other frequencies being treated as noise. The mean and confidence interval of the SNR were computed based on the full Laplace-transformed signal for the training and test sets separately, both before and after bandpass-filtering.

*Main analysis: Linear mixed-effects model for accuracy*

This statistical analysis aimed to test the accuracy of the instantaneous phase estimate across datasets, preprocessing methods, and phase estimates. We fit a linear mixed-effects model with fixed variables (1) method, (2) dataset, and (3) estimated phase bin. (1) Method is a categorical variable with nine levels (PEAP, PhastPadding, Phastimate with no padding/PEAP/PhastPadding, ETP with no padding/PEAP/PhastPadding, SSPE). (2) Dataset is also categorical with four levels (young control, elderly control, stroke affected hemisphere, stroke intact hemisphere). (3) Phase estimate was discretized into eight equally sized bins, centered around the peak. We treated the variable as categorical to allow for a non-sinusoidal fit.

Since methods may perform differently across datasets or across phase bins, we added the interaction terms method*dataset and method*phase bin. Since the stroke datasets for the affected and intact hemisphere came from the same subjects, we accounted for the repeated measures by means of a nested random effect for each subject within each dataset.

As a dependent variable, we used the phase estimation accuracy, which we calculated as the absolute normalized phase estimation error (continuous, range [0, 1]). The normalized phase estimation error (i.e., bias) was calculated as the circular distance between ground truth phase estimate and each estimator's phase estimate, normalized by pi (range [-1, 1]).

Based on the linear mixed-effects model, we assessed the impact of each predictor on accuracy using an ANOVA. We calculated marginal medians of accuracy and errors and their median absolute deviation (MAD) for all methods. To find whether Phastimate and ETP are improved by either padding method, we computed F-tests to evaluate contrasts between no padding and padding. We then compared PEAP, PhastPadding, ETP, Phastimate, SSPE, and combinations with padding algorithms that were shown to improve accuracy as determined by F-tests results. Multiple comparisons were corrected through the Benjamini-Hochberg method (Benjamini & Hochberg, 1995). We plotted a boxplot for each method-dataset combination (Fig. 3A) for the subset of methods included in multiple comparisons.

To illustrate phase estimation performance, we plotted the unfiltered data (after Laplace transform, z-scored) for the bins trough, rising, peak, and falling, as estimated by each method (Fig. 3C). These plots give insight into the behavior of each method across phases.

*Forecasting accuracy*

Besides the accuracy of the estimate for instantaneous phase, we also investigated the accuracy of estimates across time. For the predictors that allow forecasting (SSPE and Hilbert without padding were therefore not considered), we computed the median phase estimation accuracy with a 95% bootstrapping confidence interval (CI) for -100 ms pre-marker to 50 ms post-marker. The relationship was plotted for each method (Fig. 3B).

*Frequency, accuracy, biases across phase estimates*
To investigate how skewed the distribution of counts among estimated instantaneous phases is, we compared the distribution of the ground truth to the distribution resulting from each estimator. We used a circular Kuiper test (Berens, 2009) to test whether both data samples follow the same distribution. We considered any $p < 0.05$ to fail the test, i.e. both samples are likely to come from separate distributions, implying systematic biases in how common each phase is. For each predictor and ground truth, we plotted a histogram of the counts in eight bins in comparison to the ground truth distribution (Fig. 4, top row).

To test differences in accuracy between phase estimates, a sliding window (width ¼ pi, equal to bin width) was used to calculate median and 95% CI (bootstrapping) accuracy for each phase estimate. The results were plotted in comparison to the total median accuracy in Fig. 4, middle row.

In the same manner, we calculated the median bias scores with 95 % CI. We plotted the results as relative deviation from 0 (Fig. 4, bottom row).

*Correlation with SNR*
We computed the SNR of the raw EEG per trial in the frequency band (9-13 Hz) as described in the section *Quality control*. We then correlated the SNR to the root mean squared error (RMSE) at t = -1 ms of each subject to see how signal quality impacts phase estimation.

## Results

*Quality control*
The mean SNR before bandpass-filtering was -24.58 dB (95% CI: [-26.22, -22.94]) for the test set and -24.59 dB (95% CI: [-26.18, -23.01]) for the training set. After bandpass filtering, the SNR was 8.27 dB (95% CI: [7.72, 8.83]) for the test set and 8.53 dB (95% CI: [7.94, 9.13]) for the training set. Fig. 2 shows the mean power frequency spectrum for the training- and test set (see supplementary Fig. 1 for dataset-specific power-frequency spectra).

Due to noisy data, 3.72% of trials were rejected in the training set and 3.71% of trials in the test set.

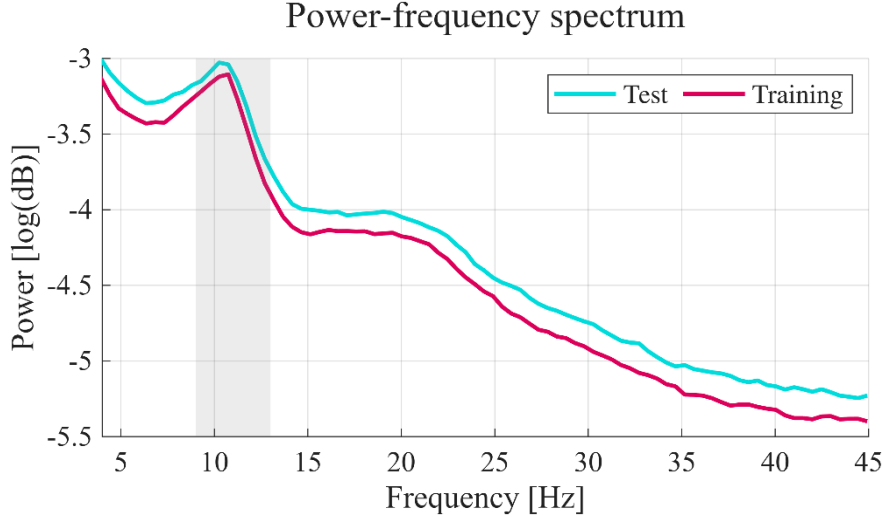

**Figure 2: Mean power-frequency spectra** of the raw Laplace-transformed data across all four data sub-sets. Power was log-transformed. The cyan line shows the test data, the magenta line shows the training data. The grey box marks the frequency band of the mu-rhythm used in this paper. Power-frequency spectra for each data sub-set can be found in supplementary figure 1.

*Main analysis - Accuracy across approaches, datasets, and bins*

For our main analysis, we fitted a linear mixed-effects model to predict accuracy from the fixed effects of dataset, method, and phase bin. Since the accuracy of processing methods may vary between the datasets and between phase bins, we included interaction terms for dataset*method, as well as phase bin*method.

The model revealed significant main effects of method ($F(8, 84,600) = 12.15$, $p < .001$) and phase bin ($F(7, 84,600) = 2.60$, $p = .011$), but not of dataset ($F(3, 84,600) = 0.51$, $p = .68$). While the interaction effect of method and phase bin was significant ($F(56, 84,600) = 1.88$, $p < .001$), there was no significant interaction between dataset and method ($F(24, 84,600) = 0.99$, $p = .48$). Thus, the phase estimation accuracy depended on the processing method and which phase bin was predicted. The relationship of phase bins and accuracy was moderated by the selected method. This accuracy did not differ between clinical and control datasets.

**Table 1: Linear mixed-effects model: prediction of accuracy from method, phase bin, and dataset.** Significant effects are marked in bold.

| Term | F | df1 | df2 | p |
|---|---|---|---|---|
| **(Intercept)** | **55.761** | **1** | **84,600** | **< .001** |
| **method** | **12.145** | **8** | **84,600** | **< .001** |
| **phase bin** | **2.605** | **7** | **84,600** | **.011** |
| dataset | 0.508 | 3 | 84,600 | .677 |
| **method × phase bin** | **1.878** | **56** | **84,600** | **< .001** |
| method × dataset | 0.986 | 24 | 84,600 | .481 |

To compare the differences between methods, a multiple comparison analysis was performed with multiple testing correction (Benjamini-Hochberg). We first tested whether Phastimate or ETP were improved by padding. Since padding ETP with PEAP significantly improved accuracy, it was included for further analyses. Consequently, the comparison included default PEAP, PhastPadding, Phastimate, SSPE, ETP, and combined PEAP + ETP.

Computing marginal medians and their median absolute deviations, PEAP had the highest accuracy (Median (Mdn) = 0.846, median absolute deviation (MAD) = 0.174), followed by PhastPadding (Mdn = 0.814, MAD = 0.194), PEAP + ETP (Mdn = 0.780, MAD = 0.217), SSPE (Mdn = 0.779, MAD = 0.206), Phastimate (Mdn = 0.776, MAD = 0.205), and ETP (Mdn = 0.754, MAD = 0.222). PEAP demonstrated significantly increased accuracy compared to all other methods except PhastPadding (see Fig. 3A, Table 2 for all comparisons, Table 3 for marginal accuracies).

**Table 2: Contrasting accuracy between methods.** Significant effects are marked in bold. P-values are Benjamini-Hochberg corrected.

| | Method 1 | Method 2 | corr. p-value | F | df1 | df2 |
|---|---|---|---|---|---|---|
| Padding VS no padding | **ETP** | **PEAP + ETP** | **0.012** | 7.414 | 1 | 84600 |
| | ETP | PhastPadding + ETP | 0.305 | 1.398 | 1 | 84600 |
| | Phastimate | PEAP + Phastimate | 0.41 | 0.747 | 1 | 84600 |
| | Phastimate | PhastPadding + Phastimate | 0.305 | 1.49 | 1 | 84600 |
| Compare methods | PEAP | PhastPadding | 0.233 | 2.017 | 1 | 84600 |
| | **PEAP** | **PEAP + ETP** | **< 0.001** | 30.75 | 1 | 84600 |
| | **PEAP** | **SSPE** | **< 0.001** | 20.706 | 1 | 84600 |
| | **PEAP** | **Phastimate** | **< 0.001** | 29.141 | 1 | 84600 |
| | **PEAP** | **ETP** | **< 0.001** | 63.43 | 1 | 84600 |
| | **PhastPadding** | **PEAP + ETP** | **< 0.001** | 14.674 | 1 | 84600 |
| | **PhastPadding** | **SSPE** | **0.007** | 8.471 | 1 | 84600 |
| | **PhastPadding** | **Phastimate** | **< 0.001** | 14.382 | 1 | 84600 |
| | **PhastPadding** | **ETP** | **< 0.001** | 38.806 | 1 | 84600 |
| | PEAP + ETP | SSPE | 0.399 | 0.858 | 1 | 84600 |
| | PEAP + ETP | Phastimate | 0.868 | 0.027 | 1 | 84600 |
| | SSPE | Phastimate | 0.359 | 1.079 | 1 | 84600 |
| | **SSPE** | **ETP** | **0.001** | 12.729 | 1 | 84600 |
| | **Phastimate** | **ETP** | **0.025** | 5.874 | 1 | 84600 |

To gain further insights into each model's characteristics, we used the model's instantaneous phase estimates at target time (t = -1 ms) to group the raw data into four phase bins. Fig. 3C shows the mean ± 95 % CI for each grouping. In this illustration, higher amplitudes indicate greater uniformity, and time shifts around t = 0 reflect systematic biases.

While all methods could discriminate between groups, the clarity of discrimination varied across methods and bins. At the target time, PEAP showed strong separation with no clear time shifts. By contrast, PhastPadding exhibited pronounced time shifts, with falling and rising data being close to peak and trough, respectively.

Whereas the ground truth, PEAP, and SSPE showed their highest discrimination at around t = 0, the other methods performed best between t = -100 to t = -50 ms, reflecting these methods' reliance on extrapolation. All methods showed reduced accuracy for t ≥ 0, highlighting the inherent challenge of forward prediction.

We then extended our analysis from the instantaneous phase estimation accuracy at t = -1 ms to phase estimation accuracy across time for the algorithms that do forecasting. Fig. 3B shows the median accuracy across time for each method. The figure demonstrates that across methods, accuracy decreased systematically from -100 ms onwards and did not reach a floor level until 50 ms into the future. At any given time, PEAP shows the highest accuracy, followed by PhastPadding. The illustration also shows how adding PEAP to ETP increased accuracy and how it shifted the timepoint of strong accuracy loss closer to 0 ms, i.e., the time of the TMS pulse.

**Table 3: Median and median absolute deviation (MAD) of accuracy and errors at t = -1 ms.** Accuracy and errors in percent.

| Method | median accuracy | MAD accuracy | median error | MAD error |
|---|---|---|---|---|
| PEAP | 84.59 | 17.39 | -0.19 | 23.14 |
| PhastPadding | 81.37 | 19.36 | 1.48 | 26.69 |
| PEAP + ETP | 78.03 | 21.74 | 2.32 | 28.22 |
| SSPE | 77.89 | 20.61 | 3.09 | 29.22 |
| Phastimate | 77.58 | 20.45 | 3.48 | 30.47 |
| ETP | 75.38 | 22.18 | 1.93 | 29.76 |

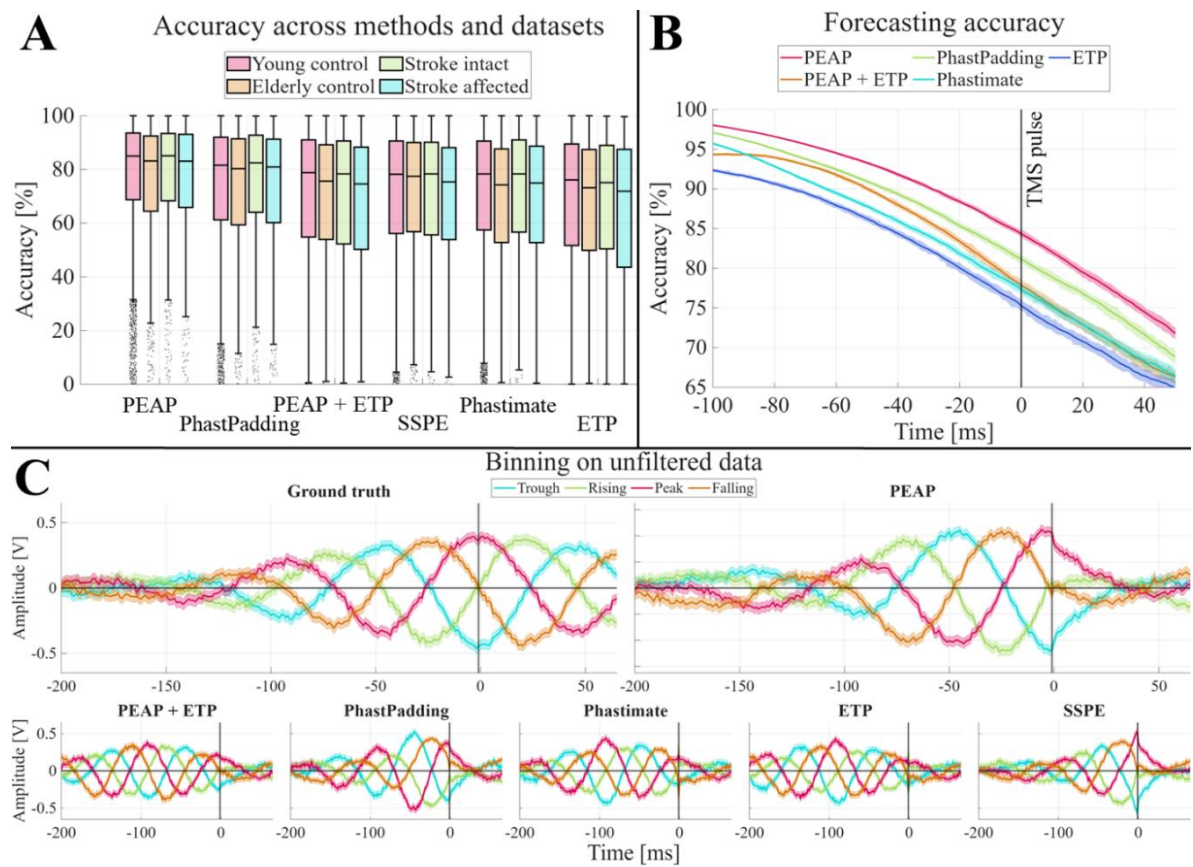

**Figure 3: Accuracy of phase estimates across methods, datasets, and time.** Accuracy shows the inverse absolute circular error scores between each method's phase estimate and ground truth phase, normalized by pi: $1 - (|\Phi - \hat{\Phi}|) / \pi$.

**A. Accuracy across methods and datasets** for instantaneous phase estimate (t = -1 ms). Methods were ranked by median accuracy. Datasets are indicated by color. A linear mixed-effects model showed significant differences (p < .001) between methods, but not between datasets. See Table 2 for multiple comparisons between methods.

**B. Forecasting accuracy.** Colored lines show accuracy of each method over time ± 95% CI. For t ≥ 0, phases were forecasted.

**C. Raw data binned into trough, rising, peak, and falling phase through each method's phase estimate at t = -1.** Detrended, Laplace-transformed, and normalized (units are thus arbitrary) data was binned by phase estimate. Each line and shaded area represent the mean ± 95% CI of the data in each bin, indicated by color. Bin width was pi/8, equivalent to the binning in the statistical analyses. The straight line indicates the target time (t = -1 ms). High amplitudes indicate strong separation of phase bins due to little contamination by incorrectly classified trials. Temporal deviations of the binning target from 0 (e.g., if the peak of the "peak" bin is located prior to 0, see ETP) indicate systematic biases.

*Differences in phase estimates between phase bins*

Our main analysis showed that phase estimation accuracy depends on the phase bin, modulated by the method used. We thus investigated the counts, accuracy, and biases across phase values. To test whether methods predict some phases systematically more often than others, we performed a circular Kuiper test (Berens, 2009) , comparing the distributions of each method to that of the ground truth. For comparison, we plotted the histogram of each method's phase estimate compared to the ground truth and the expected uniform distribution (Fig. 4, upper row). We found that all methods except PEAP show significant deviations from the ground truth distribution. Thus, PEAP was the only method that did not systematically over- or underestimate certain phase counts. PhastPadding showed the strongest deviation in an M-shape, with rising and falling phase most commonly estimated, falling being more than twice as common as trough. A smoother version of the M-shape was found for Phastimate, while SSPE expressed a similar distribution, but skewed towards early rising and early falling. ETP showed one peak at rising, which decreased towards the late falling phase. Combined with PEAP, this distribution was flattened with the highest count at trough and the lowest at late falling. PEAP only showed a slight deviation from ground truth but also displays a soft version of the M-shape. Overall, PEAP strongly reduces the systematic bias in phase count that all established methods display.

We also tested accuracy across phases with a sliding window median ± 95% CI (Fig. 4, middle row). Again, the strongest deviations from the total median accuracy were found for PhastPadding, where the trials at rising phase showed significantly decreased accuracy, 5% lower than the highest accuracy at late rising. Overall, all methods showed significant deviations from the median accuracy. Phastimate and PEAP showed a relatively stable estimate, with PEAP showing a comparably low confidence interval. This analysis indicated that all tested algorithms show some relationship between the estimated phase and the accuracy of the estimate.

Finally, we tested how the systematic bias evolves across phase estimates. Using the same sliding window approach on the median error, we determined whether estimated phases are systematically shifted to either the positive or negative direction, compared to the ground truth. A positive value represents an overestimation in phase angle estimation, while a negative value reflects that the estimated phase has an underestimated phase angle. PEAP showed no significant deviations from 0. PhastPadding and ETP showed strong biases with an overall positive trend. PhastPadding, Phastimate, and SSPE all showed that the higher counts at the rising / falling phase result from surrounding phases spilling into these bins. For ETP, mainly lower phases were spilling into the rising phase. Thus, a trough trial may be estimated by ETP to be an early rising trial, Phastimate on the other hand may categorize it into late falling. This implies that data points within the same estimated phase bin correspond to different ground truth phases and accuracy levels across methods.

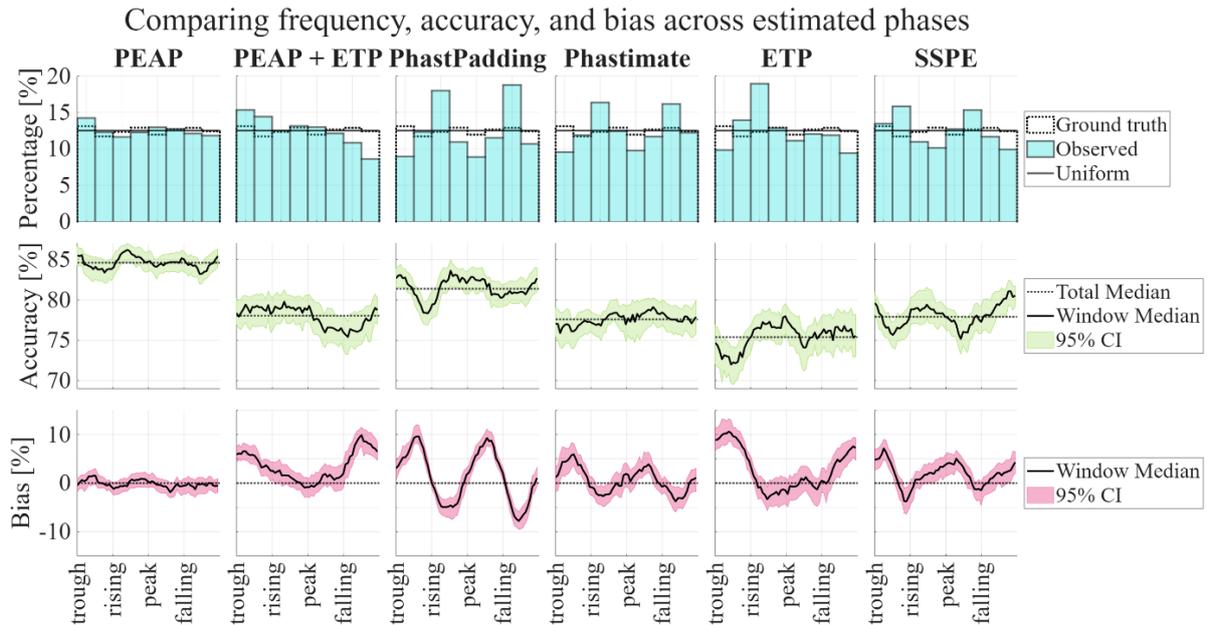

**Figure 4. Comparison of frequency, accuracy, and bias across estimated phases across methods**

**Top row: Comparison of frequency between phase bins.** Black horizontal line shows the expected uniform distribution of all phase bins occurring equally often. The dotted line shows the ground truth distribution, cyan bars show the observed distributions. PEAP is the only method where ground truth and observed distribution could be sampled from the same distribution (Kuiper test, p = 1).

**Middle row: Phase estimation accuracy across estimated phase values.** Black lines show the window median accuracy (window width: pi/4, i.e., bin width) and 95 % CI (colored areas), dashed horizontal lines show the total median accuracy.

**Bottom row: Phase estimation bias across estimated phase values.** As in the middle row, black lines and colored areas show the median and 95 % CI of normalized error scores. Dashed horizontal lines show the optimal value of no bias. Positive values mean that the estimates in the respective window are systematically tending towards higher phase angles, negative values show that estimates are skewed towards lower phase angles.

*The relationship of accuracy and bias with SNR*

To investigate the impact of the SNR on the performance of the methods, we correlated single-trial SNR to single-trial accuracy and errors. We found that for all methods, accuracy significantly improved with better SNR (all p < 0.001). The strongest correlation was $\rho = 0.2$ for Phastimate, while SSPE did not change strongly with SNR. For biases, no method showed a significant correlation between error scores and SNR. See supplementary Fig. 2 for all correlation between SNR and performances of all methods.

**Discussion**

In this study, we demonstrated that established phase estimation methods exhibit inaccuracies and systematic biases. We thus developed PEAP, a novel method that increases accuracy and substantially reduces biases. PEAP enhanced performance not only for the instantaneous phase, but also for phase forecasting, rendering it theoretically suitable for both offline and real-time applications, as previous work shows the general suitability of autoregression and padding in real-time (Chang et al., 2025; Zrenner et al., 2020). By avoiding the inherent over- or underestimation in established algorithms, PEAP provided more accurate phase estimates, which is critical for ensuring the validity and comparability of future research.

Our findings on the accuracy of different methods replicate findings from previous papers. Similar to the study by Chang (2025), we showed a significantly improved accuracy for PhastPadding compared to ETP and Phastimate. Among the tested methods, only PEAP showed higher accuracies. However, we could not replicate that combining PhastPadding with ETP or Phastimate would significantly improve either method.
Wodeyar (2021) showed higher accuracy for SSPE than Phastimate. Our work replicated this trend, but the difference is insignificant. However, our epoch length was at the lower border of potential epoch lengths for this method. Bigger epochs may improve accuracy, but given the typically short inter-trial intervals in TMS-EEG experiments, this may often be unfeasible. Contrary to Shirinpour et al. (2020), we found Phastimate to perform significantly better than ETP. A methodological difference is that they based their hyperparameters for Phastimate on previous research (Zrenner et al., 2018), while we used individualized optimized parameters.

The inaccuracies we identified are a combination of random and systematic errors, which both have substantial impact on analysis results. Random errors, i.e., noise that does not systematically distort the data, increases the variability of the data, thus weakening the relationship of phase with other variables. Higher random errors mean that effect sizes decrease, more samples are required, and relationships between variables may be discarded as false negatives (Taylor, 2022). Thus, PEAP may uncover relationships of other variables with the instantaneous phase that may not be found with other methods. Systematic errors or biases consistently skew data and can cause faulty conclusions (Taylor, 2022). For example, Phastimate and ETP may give different optimal phases for TMS stimulation. If, for example, the trough is reported to be optimal based on Phastimate, the same analysis using ETP may report the late falling phase to be best. Additionally, our results showed that established methods estimate some phases significantly more often than others. Statistically speaking, this poses issues since it violates the assumption of independent, identically distributed sampling. As samples are not identically distributed to the population, the samples are not a representative sub-sample. Moreover, the assumption of the underlying uniform probability distribution may be violated, distorting the model fit and decreasing generalizability (Blum et al., 1958). To summarize, our paper finds inaccuracies and biases that are very important to consider during method selection, statistical analysis and interpretation of results across methods.

The biases identified in our paper can be explained methodologically. We want to provide insights into three main issues: Filtering edge artifacts, edge interpolation, and model design. Filtering edge artifacts could be seen for Phastimate and PhastPadding, where they cause an M-shaped distribution of phase bin counts (Fig. 4) and biases across phase values. Since for bandpass filtering, the edge resembles a step-like function, Gibbs phenomenon occurs (Pinsky, 2023; Wilbraham, 1848). This causes samples at the signal ends to display mostly rising or falling phases. Over-estimating these phase bins is further attenuated by the fast Fourier transform that the Hilbert transform uses (Chen et al., 2011). The impact of this artifact can be reduced by increasing the size of the edge that is cut, but this increases random errors. Another option is to avoid filtering, but SSPE still displayed similar behavior, which may be caused by internal oscillatory assumptions (Wodeyar et al., 2021). We showed here that biases are even attenuated using padding based on filtered data. However, PEAP's padding with unfiltered data can greatly improve the results. Overall, filtering artifacts cause systematic distortion towards the rising and falling phase, but PEAP managed to reduce this artifact.

Cutting the edge of the filtered data is a simple solution to the edge artifact and is commonly done in the analysis of continuous data. However, most EEG research is not focused on the samples at the edge. In EEG-TMS research, the data that was cut must thus be re-created, like in Phastimate or ETP (Shirinpour et al., 2020; Zrenner et al., 2018). This leads to a difficult trade-off: Cutting a bigger edge leads to a better removal of the artifact but also causes the need to interpolate more data. Interpolating bigger amounts of data increases the insecurity of the estimate. Thus, our alternative PEAP aims to avoid cutting edges.

A third issue is the trade-off between model accuracy and interpretability. ETP offers a very intuitive method that can easily be applied and interpreted (Shirinpour et al., 2020). Understanding Phastimate and SSPE does require some methodological knowledge on machine learning. PhastPadding adds another layer of complexity by using Phastimate's output for padding without a smooth transition (Chang et al., 2025), which makes the interpretation of resulting signal analysis results quite difficult. The two-fold filtering process used causes the artifacts from Phastimate to accumulate with signal mismatches between unfiltered data and the padding data. While ETP is simple and understandable, it may oversimplify the signal. Autoregressive models and state-space models are more complex, but are able to adapt to more complex signal characteristics and signal non-stationarities (Wodeyar et al., 2021; Zrenner et al., 2020). We aimed for a good tradeoff of accuracy and complexity, where the autoregressive model is interpretable and the interpolated data can be inspected visually, but the model can still change over time and adapt to the current samples.

Applying methods in research requires them not only to be unbiased, accurate, and interpretable, but also fast. We did not test processing speed in our research, since this is highly dependent on software engineering skills, which is not our central expertise. We do want to highlight here that this is a crucial aspect for real-time applications since forecasting accuracy decreases with time (Fig. 3B). We thus encourage further research on software implementation that is focused on processing time.

In this paper, we applied PEAP to a very simple pre-processing pipeline. For more complex analyses, we would advise padding the data prior to processing steps that can distort the data. For example, padding should be applied prior to high-pass filtering for ICA. Future research may aim to compare the algorithms presented in this paper in a variety of standard data processing pipelines.

Our research gives an overview on the behavior of different phase estimation algorithms for healthy and stroke data. It should be noted that the stroke data was collected in a research context with high standards for signal quality. Purely clinical data may not reflect the same SNRs, limiting the generalizability of our findings to clinical applications.
Additionally, we conducted our analyses offline. In future work, we will validate our algorithm online and expect its behavior to reflect our findings here. Additionally, future work might look into creating a toolbox that lets the user pick the optimal tool for their use and data. Given that different algorithms show different strengths, a consensus tool could also be developed.

## Conclusion

In this paper, we present Phase Estimation through Autoregressive Padding (PEAP) as a novel method to estimate the instantaneous phase of oscillatory signals that overcomes the limitations of established algorithms. Through an overview of accuracy and biases of different methods, we provide insights for the comparison of results. By including data of stroke patients, we show that the implementation of all presented algorithms is principally feasible for clinical use.

## Code availability

The code used for this publication as well as code and documentation for using PEAP can be found on github: MiriamKirchhoff/PEAP.

## Author contributions

Miriam Kirchhoff: Conceptualization, Methodology, Software, Validation, Formal analysis, Writing - Original draft, Visualization, Project administration
Johanna Rösch: Conceptualization, Data curation, Writing - Review & Editing
Maria Ermolova: Data curation
Oskari Ahola: Data curation, Writing - Review & Editing
Sarah Harders: Data curation
Juliana Hougland: Conceptualization, Writing - Review & Editing
Ulf Ziemann: Supervision, Funding acquisition, Writing - Review & Editing

## Funding
This study is part of the ConnectToBrain project that has received funding from the European Research Council (ERC) under the European Union's Horizon 2020 research and innovation programme (grant agreement No. 810377).

## Acknowledgements
The authors declare that they have no conflicts of interest.

**Supplementary materials**

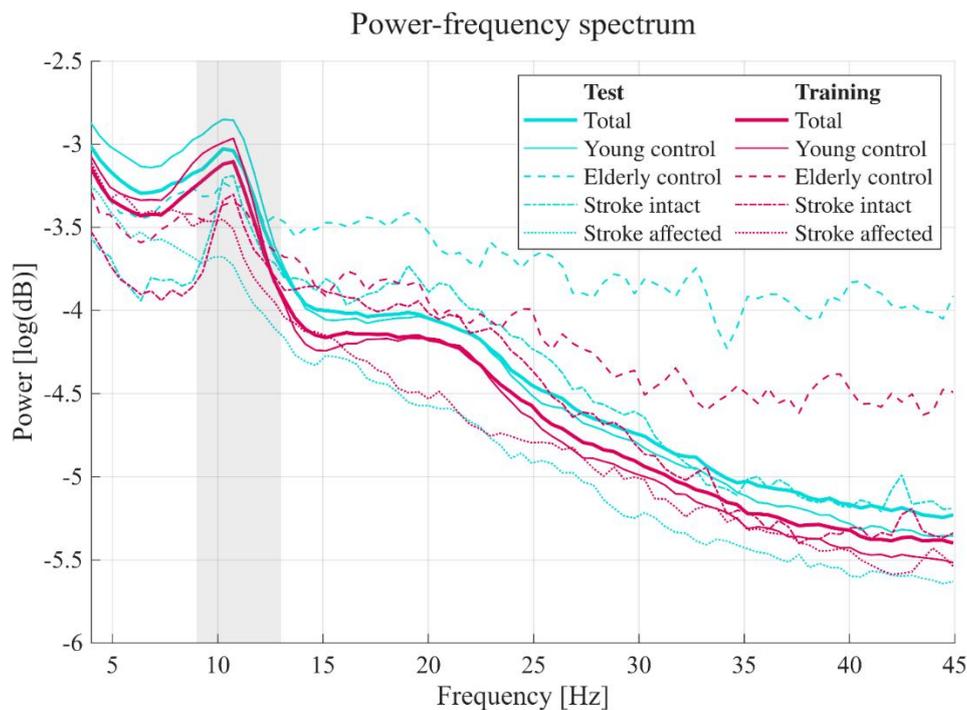

***Supplementary Figure 1: Mean power-frequency spectra across datasets** of the raw Laplace-transformed data across subjects. Power was log-transformed. The cyan lines show the test data, the magenta lines show the training data. The thick, continuous lines show the total mean for each group. The grey box marks the mu-rhythm frequency used in this paper.*

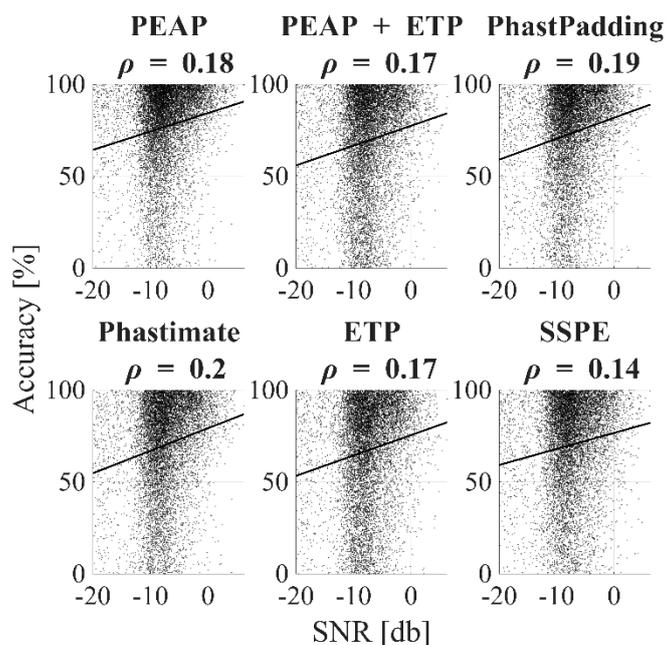

***Supplementary Figure 2: Correlation of accuracy with single-trial SNR.** Correlation coefficients ρ are indicated for all methods, p < 0.001 for all correlations. Black lines show the linear regression fit.*